# SOUND LOCALIZATION USING COMPRESSIVE SENSING

Hong Jiang, Boyd Mathews and Paul Wilford

*Bell Labs, Alcatel-Lucent, 700 Mountain Ave, Murray Hill, NJ 07974, USA*
*{jiang, btmathews, paw}@alcatel-lucent.com*

**Abstract:**      In a sensor network with remote sensor devices, it is important to have a method that can accurately localize a sound event with a small amount of data transmitted from the sensors. In this paper, we propose a novel method for localization of a sound source using compressive sensing. Instead of sampling a large amount of data at the Nyquist sampling rate in time domain, the acoustic sensors take compressive measurements integrated in time. The compressive measurements can be used to accurately compute the location of a sound source.

## 1 INTRODUCTION

This paper is concerned with localization of sound source in a network of distributed acoustic sensors. In a typical application of wireless distributed sensing network for surveillance, acoustic sensors are distributed remotely in a region of interest. Each sensor collects samples of sound wave arrived at its location from a sound source. The sensors transmit the collected samples to a processing center where the samples are analyzed and the location of the sound source is estimated and tracked. The sensors are small, low-cost devices that may be powered by batteries. Each sensing device integrates the functionality of converting acoustic pressure to an electronic signal, sampling the electronic signal and transmitting the samples via wireless communications. In such a scenario, it is important for the sensors to be built with high reliability, low complexity and low power consumption. The best way to achieve all these is to reduce the sampling rate required at the sensing devices. A low sampling rate at the sensors not only reduces the complexity of the devices but also significantly reduces power consumption of the circuits used for both analog to digital conversion (ADC) and the wireless transmission. A low rate wireless transmission also increases reliability of the data transmission.

The Nyquist sampling rate of the acoustic sensing devices is determined by the frequency range that is needed in an application. For localization of sound sources, the sampling rate needs to be high enough to cover the acoustic frequency range that is necessary to perform localization with a prescribed resolution and accuracy. Most sound localization techniques are based upon determining the difference in propagation time for a signal emitted from a source to arrive at two or more sensors. This is known as the time-difference-of-arrival (TDOA). In these techniques, sound waves are sampled at the Nyquist sampling rate by the sensors, and then TDOA is determined from the collected samples. The cross-correlation method (Knapp and Carter, 1976, Valin, Michaud, Rouat and L´etourneau, 2003) uses the location of the peak value in the sample cross-correlation between two sensors to estimate TDOA. Other methods (Benesty, Chen and Huang, 2008) make use of channel response functions of the signals arrived at the sensors.

Compressive sensing (Candès, Romberg and Tao, 2006) is an emerging theory for representing and reconstructing sparse signals by using far fewer measurements than the number of Nyquist samples. When a signal has a sparse representation, the signal may be reconstructed from a small number of measurements from linear projections onto some basis. Furthermore, the reconstruction has high probability of success even if a random sensing matrix is used.

In this paper, we develop a method for sound localization in a distributed sensing network using compressive sensing. In order for localization to be reliable for a variety of sound events, we don't make the assumption that the sound wave from a sound source is sparse. Instead, we observe that in most reasonable circumstances, the acoustic signal at one sensor has a sparse representation if the acoustic signal at another sensor is known. We regard the signal at one sensor to be the output of a linear system with the input as the signal at another sensor, and we assume that the linear system can be approximated by a finite impulse response filter with a very small number of nonzero coefficients. Consequently, it is possible for sensors to take and transmit very small number of measurements if the signal at one of the sensors is known.

By using compressive measurements, a sound sensor is only required to make and transmit samples at a low rate, which improves the reliability of the sensor and reduces the power assumption. Furthermore, since the measurements are made by linear projections, the complexity of acquiring the measurements is also low.

The paper is organized as follows. In section 2, we describe existing sound localization techniques. Our method of localization using compressive sensing is described in Section 3. Some simulation and experiment results are presented in Section 4, and the conclusion is provided in Section 5.

# 2 SOUND LOCALIZATION TECHNIQUES

Figure **1** illustrates a distributed sensor network. The sensors are distributed with a known geometry. The sensors make samples of the sound waves, and transmit the samples to a processing center (not shown) for analysis.

## 2.1 Direction of Arrival (DOA) and Time Difference of Arrival (TDOA)

The location of the sound source can be estimated from the time differences between the arrival times of the sound wave at the sensors. For the purpose of this paper, we consider the case where the sound source is sufficiently distant so that the wavefront arriving at the array approximates a plane. Figure 1 shows the derivation of the DOA shown as angle $\theta$ between the segment $\overline{x_0 x_1}$ and the arriving sound. The quantity $d_1 cos(\theta)$ can be calculated by measuring the time delay, $\tau_{01}$, for the wavefront to propagate from $x_0$ to $x_1$. Using $c$ to denote the velocity of sound in air, we have $c\tau_{01} = d_1 \cos(\theta)$, and thus $\theta = \cos^{-1}(c\tau_{01}/d_1)$. In this way, a pair of sensors can be used to determine the relative direction to the sound source. Multiple pairs can be used to triangulate upon the source position.

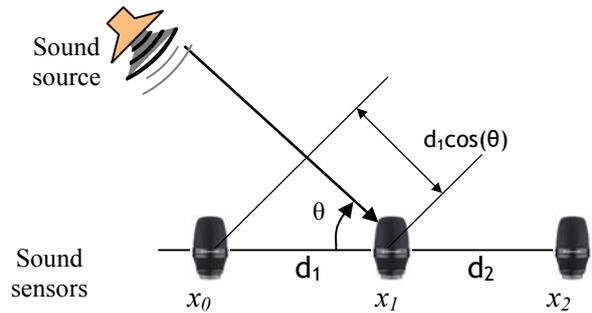

Figure 1 Distributed sensor network.

## 2.2 Cross-correlation

Cross-correlation of the sensor signals, proposed by (Knapp, Carter 1976) is the most straightforward means to measure TDOA. Given a single sound source in a quiet anechoic space, the output of each sensor is a transduction of the source signal affected only by the delay and attenuation associated with the path length between the source and sensor. The cross-correlation of two such signals is maximized at the time lag corresponding to the difference in path delays. Let $x_0(i)$ and $x_1(i)$ represent the samples of signals arriving at sensors $x_0$ and $x_1$. The samples are processed in blocks of length N. For each block of samples, we calculate the cross-correlation function

$$r_{12}(\tau) = \sum_{i=0}^{N-1} x_1(i)x_2(i-\tau),$$

where the range of $\tau$ is limited to $\tau \leq d_{max}/c$ by the maximum distance between sensors. TDOA is determined by finding the value of delay which maximizes the cross-correlation

$$\tau_{TDOA} = \arg\max_{\tau} r(\tau).$$

Several aspects of cross-correlation for TDOA estimation should be noted. First, correlation of periodic signals is also periodic. Spatial aliasing will occur for periodic signals with a wavelength shorter than twice the sensor spacing. Also, even for non-periodic signals, echoes can degrade the accuracy of this method by creating local maxima in the cross-correlation function.

The cross-correlation approach for TDOA estimation is normally performed in the frequency domain (Valin, Michaud and Rouat and L'etourneau, 2003) and many enhancements have been proposed based upon frequency weighting functions. The smoothed coherence (SCOT) method (Carter, Nuttall and Cable, 1973) attempts to minimize the influence of the source signal power. The phase transform (PHAT) method (Knapp and Carter, 1976) removes the source amplitude from the cross-spectrum calculation altogether. However, we will not elaborate further in this direction and instead consider a quite different way of looking at TDOA estimation.

### 2.3 Channel impulse response

The adaptive eigenvalue decomposition (AED) algorithm presented by (Benesty, Chen and Huang, 2008) measures TDOA between two sensors by estimating the channel response from the source to each sensor. The direct path feature is extracted from each channel response and the timing difference between these two features is taken as the TDOA estimate.

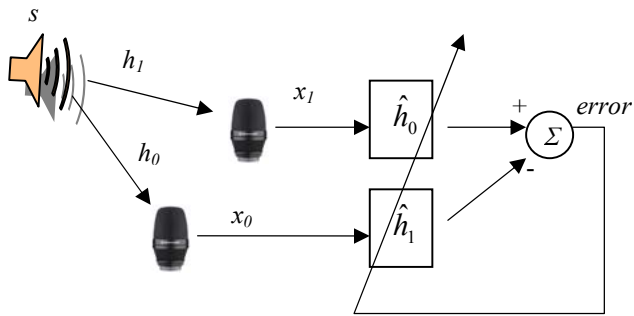

Figure 2 Adaptive estimate of channel responses.

As illustrated in Figure 2, given the source signal, $s$, is not known, the fundamental observation allowing estimation of the two channel responses is

$$x_1^T * h_0 - x_0^T * h_1 = 0 .$$

A constrained LMS algorithm, presented in (Benesty, Chen and Huang, 2008), can be formulated allowing adaptation towards the concatenated channel response estimates $\hat{h}_0$ and $\hat{h}_1$. After convergence, the maximum amplitude of each response is assumed to correspond to the arrival of the direct sound and the time difference between these peaks is the TDOA. A significant advantage of AED over cross-correlation is its robustness under reverberant conditions.

## 3 LOCALIZATION USING COMPRESSIVE SENSING

### 3.1 Compressive sensing background

A brief review of compressive sensing is given in this subsection. A signal $x \in \mathfrak{R}^N$ is sparse if it is comprised of only a small number of nonzero components when expressed in certain basis (Candès, Romberg and Tao, 2006). Specifically, $x$ is $S$-sparse if there exists an invertible matrix $\psi \in \mathfrak{R}^{N \times N}$ and a vector $h \in \mathfrak{R}^N$ such that

$$x = \psi h , \text{ and } \|h\|_0 = S << N . \qquad (3.1)$$

In (3.1), $\|h\|_0$ is the number of nonzero elements of $h$.

Since $h$ has $S$ nonzero elements, signal $x$ can be uniquely represented by no more than $2S$ numbers in a straightforward way: the locations and the values of the nonzero elements of $h$. However, this representation requires the availability of all $N$ samples of signal $x$. In other words, this representation still requires the signal $x$ to be acquired with $N$ samples.

Compressive sensing makes it possible to acquire a sparse signal using far fewer than $N$ measurements. In compressive sensing, a signal is projected onto a measurement basis, and the projections can be used to recover the signal. Specifically, let $\phi \in \mathfrak{R}^{M \times N}$ be a sensing matrix. Then the measurements $y \in \mathfrak{R}^M$ are given by

$$y = \phi x . \qquad (3.2)$$

The number of measurements $M$ can be much smaller than the length of vector $x$, $N$. Under the conditions that 1) $\phi$ and $\psi$ are incoherent; and 2) $M$ is large enough with respect to $S$ (Candès, Romberg and Tao, 2006), the sparse signal $x$ can be reconstructed from the measurements $y$ by solving the following minimization problem:

$$\min \| h \|_1 \text{ subject to } \phi\psi\, h = y. \qquad (3.3)$$

In (3.3), $\| h \|_1$ is the sum of the absolute values of the components of $h$. After $h$ is found from (3.3), $x$ is computed from $x = \psi\, h$. The minimization problem (3.3) can be solved by using standard linear programming techniques.

Although it is difficult to verify the incoherence condition for given sensing matrix $\phi$ and sparsity basis $\psi$, it is known (Candès, Romberg and Tao, 2006) that for a given sparsity basis $\psi$, a random sensing matrix $\phi$ has a high probability of being incoherent with $\psi$. In other words, the signal $x$ has a high probability of being recovered from random measurements. In practice, it has been found that randomly permuted rows of the Walsh-Hadamard matrix may be used to form a sensing matrix with satisfactory results (Li, Jiang and Wilford and Zhang, 2011, Jiang, et al., 2012). Sensing matrix formed from shifted maximum length sequence (MLS) will be discussed later.

### 3.2 Sound localization

We now describe our method for performing sound localization by using compressive sensing.

As shown in Figure 1, let $s(t)$ represent the acoustic signal from the sound source, and $x^{(i)}(t)$ represent the signal of the sound source arriving at the sensor $i$. Then signal $x^{(i)}(t)$ can be written as

$$
\begin{aligned}
x^{(i)}(t) &= \left( \hat{h}^{(i)} * s + \hat{\eta}^{(i)} \right)(t) \\
&= \int_0^t \hat{h}^{(i)}(\tau) s(t - \tau) d\tau + \hat{\eta}^{(i)}(t).
\end{aligned}
\qquad (3.4)
$$

In Eq. (3.4), $\hat{h}^{(i)}(t)$ is the impulse response of the channel from the sound source to sensor $i$, and $\hat{\eta}^{(i)}(t)$ is the Gaussian noise.

We assume that the channel from the sound source to sensor $i = 0$ is invertible, i.e., there is a deconvolution of $x^{(0)}(t)$ so that

$$s(t) = \left( g * x^{(0)} \right)(t) + \eta^{(0)}(t). \qquad (3.5)$$

Then Eqs (3.4) and (3.5) give rise to the following equations

$$
\begin{aligned}
x^{(i)}(t) &= \left( h^{(i)} * x^{(0)} \right)(t) + \eta^{(i)}(t), \\
&= \int_0^t h^{(i)}(\tau) x^{(0)}(t - \tau) d\tau + \eta^{(i)}(t), \\
i &= 1, 2, \dots
\end{aligned}
\qquad (3.6)
$$

where

$$
h^{(i)} = g * \hat{h}^{(i)}, \eta^{(i)} = \hat{\eta}^{(i)} - g * \hat{\eta}^{(0)}, \\
i = 1, 2, \dots
\qquad (3.7)
$$

Equations in (3.6) and (3.7) show that if a deconvolution of $x^{(0)}(t)$ exists, then each signal arriving at the other sensors $x^{(i)}(t)$, $i = 1, 2, \dots$, is a convolution of $x^{(0)}(t)$ plus noise.

Let us now consider the discrete samples of the acoustic signals with sample duration of $T$. Let $x^{(i)} \in \Re^N$, $x_n^{(i)}$, $n = 0, \dots, N$ be the samples of the signal $x^{(i)}(t)$, and $h^{(i)} \in \Re^N$, $h_n^{(i)}, n = 0, \dots, N$ be the samples of $h^{(i)}(t)$. For convenience, although it is not necessarily, we assume that $N$ is an even number.

We assume that the number of samples $N$ is large enough so that the support of $h^{(i)}(t)$ is contained within the interval $[0, NT]$, that i.e.,

$$h^{(i)}(t) = 0, t \notin [0, NT], i = 1, 2, \dots \qquad (3.8)$$

Eq. (3.8) may not be satisfied for any finite $N$ if the signal at sensor $i = 0$ contains echoes. This is because even though $\hat{h}^{(i)}(t)$ and $\hat{h}^{(0)}(t)$ may have a finite support, the deconvolution $g(t)$, and hence $h^{(i)}(t) = g(t) * \hat{h}^{(i)}(t)$ may not. Nevertheless, the amplitude of $h^{(i)}(t)$ outside of the interval $[0, NT]$ can be made small enough to be ignored for sufficiently large $N$ so that it is reasonable to assume Eq. (3.8)

holds in practice for large $N$.

The discretized version of Eq. (3.6) becomes

$$x^{(i)} = \psi_0 h^{(i)} + \eta^{(i)}, \; i = 1,2,\ldots \qquad (3.9)$$

where

$$x^{(i)} = \begin{bmatrix} x_0^{(i)} \\ \vdots \\ x_N^{(i)} \end{bmatrix}, h^{(i)} = \begin{bmatrix} h_0^{(i)} \\ \vdots \\ h_N^{(i)} \end{bmatrix}, \qquad (3.10)$$

$$\psi_0 = \begin{bmatrix} x_{\frac{N}{2}}^{(0)} & \cdots & x_0^{(0)} & \cdots & x_{\frac{N}{2}}^{(0)} \\ \vdots & \ddots & \vdots & \ddots & \vdots \\ x_{\frac{N}{2}}^{(0)} & \cdots & x_N^{(0)} & \cdots & x_{\frac{3N}{2}}^{(0)} \end{bmatrix}.$$

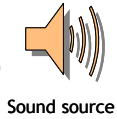

Sound source

The vectors $h^{(i)}$, $i = 1,2,\ldots$, are sparse because most of its entries are zero or small. The entries of $h^{(i)}$ with largest absolute values provides information on the time delay between signals $x^{(i)}(t)$ and $x^{(0)}(t)$.

For example, if the time delay between $x^{(i)}(t)$ and $x^{(0)}(t)$ is an exact integer multiple of the sample duration $T$, then the time delay between the two signals is given by

$$\Delta t^{(i)} = \left( \arg \max_j \{|h_j^{(i)}|\} - \frac{N}{2} \right) T. \qquad (3.11)$$

Time delay of a fraction of sample duration may be obtained by interpolation using a few neighboring values of the entry with maximum absolute value.

Eq. (3.9) shows that $x^{(i)}$ is a sparse signal with sparsity basis $\psi_0$. Note that no assumption has been made regarding the sparsity of the acoustic source signal $s(t)$. Regardless of whether or not the source signal $s(t)$ is sparse, the signal $x^{(i)}$ sampled at sensor $i$, $i = 1,2,\ldots$, always has a sparse representation in the basis $\psi_0$ after $x^{(0)}$ is available. Therefore, the theory of compressive sensing may be directly applied to the sparse signals $x^{(i)}$, $i = 1,2,\ldots$

Let $\phi_i \in \Re^{M \times N}$ be the sensing matrix at sensor $i$. Each of the sensors $i$, $i = 1,2,\ldots$ takes measurements

$$y^{(i)} = \phi_i x^{(i)} \in \Re^M. \qquad (3.12)$$

However, sensor $i = 0$ takes samples $x^{(0)}$ of the sound wave in the traditional way using the Nyquist sample rate. This entire process is shown in Figure 3.

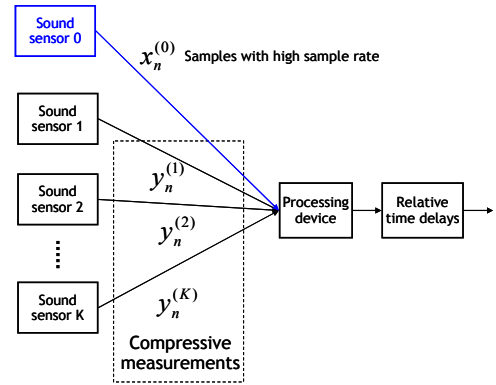

Figure 3 Distributed sensing network with compressive sensing.

In Figure 3, a network of $K + 1$ sound sensors is shown. The sensors are configured so that one sensor takes the traditional Nyquist samples, but the other $K$ sensors make compressive measurements.

In order to compute the time deference between $x^{(i)}$ and $x^{(0)}$, the Nyquist sampled signal $x^{(0)}$ is used to form the sparsity basis $\psi_0$ by using Eq. (3.10). Then the channel response $h^{(i)}$ is computed from the minimization problem

$$\min_{h^{(i)}} \|h^{(i)}\|_1, \text{ subject to } \phi_i \psi_0 h^{(i)} = y^{(i)}, \quad (3.13)$$

or, in practice

$$\min_{h^{(i)}} \left\{ \|h^{(i)}\|_1 + \frac{\mu}{2} \|\phi_i \psi_0 h^{(i)} - y^{(i)}\|_2 \right\}. \quad (3.14)$$

In (3.14), $\mu > 0$ is a constant.

### 3.3 Implementation considerations

We discuss some practical issues of implementation in this subsection.

#### Sensing matrix

The sensing matrix $\phi$ may be formed from a maximum length sequence (m-sequence). M-sequences have been traditionally used in audio processing (Rife and Vanderkooy, 1989), but they are used for a different purpose in this context. Let $p_n$, $n = 1, ..., N$ be a binary m-sequence generated from a primitive polynomial. Then each row of the sensor matrix $\phi$ is formed by a shifted sequence of $p_n$. For example the entries of the sensor matrix can be defined by

$$\phi_{ij} = 1 - 2 p_{(j+i) \bmod N},$$
$$i = 1, ..., M, j = 1, ..., N \quad (3.15)$$

The advantage of using the shifted m-sequences to form the sensor matrix is that the m-sequence can be easily implemented in hardware by using linear feedback shift registers and hence reducing the complexity of matrix generation in the sensors.

All sensors need not use the same sensing matrix. However, different sensing matrix can be created with the same m-sequence, but with different shifts for the rows. Again, this arrangement helps reducing complexity.

***Detection confidence indicator***

The solution to minimization problem (3.14) has a stochastic nature. This can be viewed from two aspects. First, when a random sensing matrix such as (3.15) is used, the compressive sensing theory only guarantees the success of recovery with a high probability. Therefore, the peak value in the solution to (3.14) only provides the correct time delay in the statistical sense. Secondly, the solution to (3.14) is only meaningful when there is a sound from the source. For example, the signals at the sensors are comprised of only noise when the source is silent, and the solution to (3.14) would result in a peak at a random location.

The stochastic nature can be exploited to our advantage to create a metric of how accurate the solution is. In other words, we are able to utilize the characteristics of compressive sensing to create an indicator on how confident we are about the detection of the sound source.

When $M$ measurements $y^{(i)}$ are received from sensor $i$, they are used in (3.14) to compute an estimate of the time delay $\Delta t^{(i)}$ as given by Eq. (3.11). Similarly, any subset of the measurements may also be used to repeat the process. Therefore, the minimization process (3.14) may be performed multiple times, each time with a randomly selected small number of measurements removed from $y^{(i)}$, to compute multiple estimates of the time delay $\Delta t^{(i)}_j$. Here the subscript $j$ denotes the repetition index of process (3.14) for the estimate of the same time delay $\Delta t^{(i)}$. The values of $\Delta t^{(i)}_j$, $j = 1, ...,$ may be processed to produce a final estimate $\Delta t^{(i)}$ and a metric of confidence $C^{(i)}$. For example, they may be defined as

$$\Delta t^{(i)} = \underset{j}{\mathrm{median}} \{\Delta t^{(i)}_j\},$$
$$C^{(i)} = \frac{1}{\max_j \{\Delta t^{(i)}_j\} - \min_j \{\Delta t^{(i)}_j\}} \quad (3.16)$$

Since these computations are done at a processing center, not at the sensors, the complexity is not a concern.

# 4 SIMULATION AND EXPERIMENT

We present some simulation and experiment results in this section.

## 4.1 Simulation with three sensors

In this simulation, the sensing network consists of three sensors as shown in Figure 4.

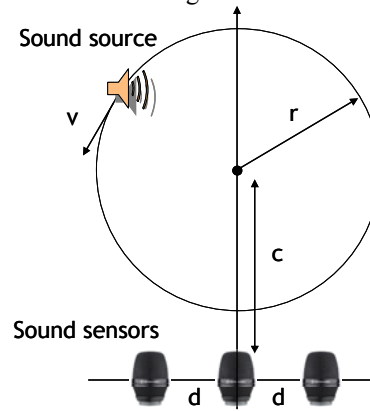

Figure 4 Moving sound source with three sensors.

The sensors are placed horizontally on the x-axis with distance $d$. A sound source is moving with speed $v$ in a circle of radius $r$ with the center on the y-axis and a distance of $c$ away from the x-axis. The sensor in the middle is chosen to take Nyquist samples, i.e., $i = 0$.

The signal source is chosen from (Lin, Lee and Saul, 2004) and given by the equation

$$s(t) = e^{-\frac{1}{2}\left(\frac{t}{\tau}\right)^2} \sin 2\pi f_0 t . \qquad (4.1)$$

The following parameters are used in for the model network of Figure 4.

$$d = 1m \qquad c = 7m \qquad r = 5m$$
$$v = 0.47 m/s \quad f_0 = 16kHz \quad \tau = 10 \sec \qquad (4.2)$$

The middle sensor, $i = 0$, takes Nyquist samples of the arriving sound signal at the sample rate of $f_s = 16kHz$. The sensors at two sides, $i = 1,2$, make compressive measurements of the arriving signals. The sensing matrix $\phi$ is formed from shifted m-sequences as described in Section III. Each measurement is a projection of $N = 4095$ samples of $f_s = 16kHz$. In other words, the estimate of time delay is performed on blocks of $N = 4095$ samples, which corresponds to a time duration of 0.256 seconds. For each block, $M = 40$ measurements are used in the minimization process (3.14). For each set of measurements from sensors $i = 1,2$, the solution to the minimization problem (3.14) produces an estimate for the time difference of arrival (TDOA) between the side sensor and the middle sensor, $\Delta t^{(i)}$, by using Eq. (3.11). The estimate is accurate up to the sample duration. The results of the time differences are shown in Figure 5.

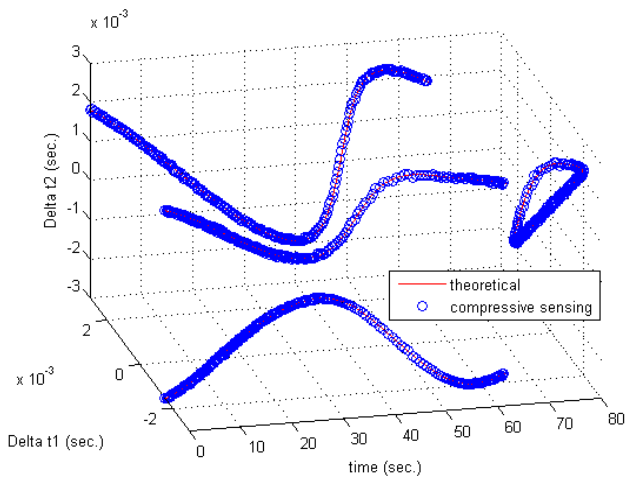

Figure 5 Time differences at the sensors for a moving sound source.

In Figure 5, the TDOA $\Delta t^{(i)}$, $i = 1,2$, are plotted at different time instances. The x-axis in Figure 5 is the time instances. The y- and z-axes are for $\Delta t^{(1)}$ and $\Delta t^{(2)}$, respectively. The sound source moves at a constant speed in a circle, and it takes about 66.3 seconds to complete the circle. As the sound source moves along the circle, the pair of TDOAs ($\Delta t^{(1)}$, $\Delta t^{(2)}$) traces out a loop in the $\Delta t^{(1)} - \Delta t^{(2)}$ plane that has a shape of a cloth hanger. The computed result is compared to the theoretical result, and the error in the computed result is within one sample duration ($62.5\mu s$) since the resolution of the estimate as given in (3.11) is the sample duration.

An important observation from this simulation is that the result was achieved by using only $M = 40$ measurements for each of the side sensors $i = 1,2$, as opposed to the traditional Nyquist samples of $N = 4095$. This represents a compression ratio of more than 100 times. In other words, each of the sensors $i = 1,2$ takes 40 measurements and transmits them to the processing center, instead of traditional 4095 samples. The compression ratio of more than 100 times implies that the sensors are able to transmit the measurements much more reliably and power-efficiently. Also the compression ratio is achieved with a very low complexity of projections with the sensing matrix formed from the shifted m-sequences.

## 4.2 Experiments with two sensors

In this subsection, we present two experiments with data from actual recordings from two microphones. In the experiments, two microphones are placed in a room about 11 cm apart. The microphones are used to make recordings of a person reading an article. Signals from both microphones are sampled at the same Nyquist sampling rate. The samples from one microphone is used to form the sparsity basis $\psi_0$, and the samples from another microphone is used to form the compressive measurements using a random sensing matrix formed by the shifted m-sequences. The number of full samples is $N = 4095$, and the number of measurements is $M = 40$, representing a compression ratio of 100. The estimate of difference in signal arrival times is computed by solving the minimization problem (3.14), and the results are compared to those obtained by using the classic cross-correlation method by using the full $N = 4095$ samples.

In the first experiment, the setup is such that the signals from the microphones are clean, and the speaker remains relatively still. In this experiment, the sample rate is $f_s = 16kHz$. The result, compared with the cross-correlation, is shown in Figure 6.

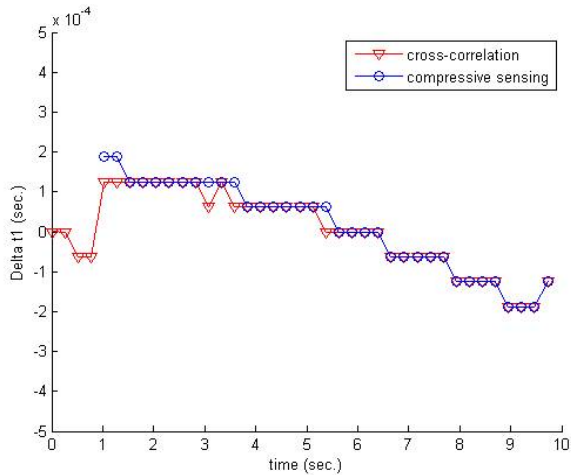

Figure 6 Two sensors with clean signals.

At the start of the recording, there is a segment of silence. As described in Section III, the compressive sensing method is able to ignore the results from that segment by using the confidence indicator (3.16). The confidence indicator is much less than 1 during the initial segment in which the cross-correlation result is unreliable as well. After the initial silent segment, the confidence indicator is infinity and the result matches with the cross-correlation very well.

In the second experiment, the setup represents a difficult environment in which echoes and noise are abundant. In addition, the speaker is constantly moving around. The sample rate for this experiment is $f_s = 44.1kHz$. The result, compared with the cross-correlation, is shown in Figure 7.

This is a difficult situation as is evident by many erroneous results from the cross-correlation method. Admittedly, the compressive sensing method made many erroneous calculations as well, but the use of the confidence indicator is able to identify and eliminate the unreliable data points. The eliminated data points from compressive sensing are not plotted in Figure 7, which can be seen from the connected line without the symbol. It is worthwhile to point out that it is possible to apply certain processing to eliminate the erroneous results from the cross-correlation method as well. However, the purpose of showing the result from the cross-correlation method without further processing is to illustrate that the environment used in this experiment is a difficult one, and even in this difficult environment, the compressive sensing method is able to produce reasonable result with a high compression ratio of more than 100 to 1.

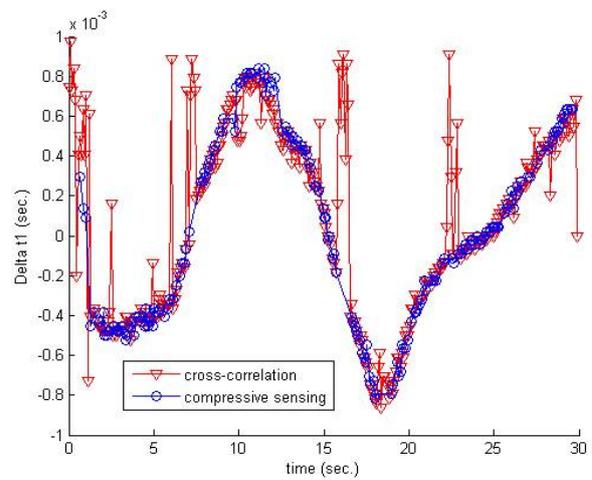

Figure 7 Two sensors with noisy signals.

## 5 CONCLUSION

Compressive sensing is an effective technique for localization of sound source in a sensing network. Compressive measurements can be used to reliably estimate the time difference of arrival (TDOA) of sound signals at the sensors, without any assumption on the sparseness of the sound source. We have demonstrated reliable detection and tracking of sound source by using compressive measurements with a compression ratio of more than 100 times, as compared to the traditional Nyquist sampling.